\documentclass[journal,comsoc,twoside]{IEEEtran}

\usepackage{amsmath}
\usepackage{color}
\usepackage{multirow}

\usepackage{amsthm} 
\usepackage{amssymb}
\usepackage{float}
\usepackage{nccmath}
\usepackage{graphicx}
\usepackage{subcaption}
\usepackage[linesnumbered,ruled]{algorithm2e}
\usepackage{booktabs}
\usepackage{mwe}
\usepackage{cite}
\usepackage{graphicx}
\usepackage{textcomp}
\usepackage[normalem]{ulem}
\usepackage{xcolor}
\usepackage{tabulary}
\usepackage{array} 

\usepackage{mathtools}

\usepackage{amsmath,amssymb,amsfonts}
\usepackage{algpseudocode}
\usepackage{xspace}
\usepackage{kotex}

\begin{document}
\title{Joint Pilot Design and Channel Estimation using Deep Residual Learning for Multi-Cell Massive MIMO under Hardware Impairments} 
%

\author{
Byungju Lim, Won Joon Yun, Joongheon Kim,~\IEEEmembership{Senior Member,~IEEE}, and Young-Chai Ko,~\IEEEmembership{Senior Member,~IEEE} 
\thanks{This work was supported by the National Research Foundation of Korea (NRF) grant funded by the Korea government (MSIT) (NRF-2020R1A4A1019628).}
\thanks{B. Lim, W. J. Yun, J. Kim, and Y. -C. Ko are with the School of Electrical Engineering, Korea University, Seoul 02841, Republic of Korea (e-mails: \{limbj93, ywjoon95, joongheon, koyc\}@korea.ac.kr)}
\thanks{Y. -C. Ko is a corresponding author of this paper.}
}

\maketitle

\begin{abstract}
In multi-cell massive MIMO systems, channel estimation is deteriorated by pilot contamination and the effects of pilot contamination become more severe due to hardware impairments.  
In this paper, we propose a joint pilot design and channel estimation based on deep residual learning in order to mitigate the effects of pilot contamination under the consideration of hardware impairments.
We first investigate a conventional linear minimum mean square error (LMMSE) based channel estimator to suppress the interference caused by pilot contamination.
After that, a deep learning based pilot design is proposed to minimize the mean square error (MSE) of LMMSE channel estimation, which is utilized to the joint pilot design and channel estimator for transfer learning approach.
For the channel estimator, we use a deep residual learning which extracts the features of interference caused by pilot contamination and eliminates them to estimate the channel information. 
Simulation results demonstrate that the proposed joint pilot design and channel estimator outperforms the conventional approach in multi-cell massive MIMO scenarios. Furthermore, the joint pilot design and channel estimator using transfer learning enhances the estimation performance by reducing the effects of pilot contamination when the prior knowledge of pilot contamination cannot be exploited.
\end{abstract}

\IEEEpeerreviewmaketitle


\section{Introduction}\label{sec:intro}
\subsection{Backgrounds and Motivation}
Massive multiple-input multiple-output (MIMO) has been attracted considerable attention in wireless communications to meet the high data rate requirements and improve the link reliability \cite{JSTSP_MIMO}. In addition, massive MIMO has the advantages of multiplexing gain, simple signal processing, and cost reduction in radio frequency (RF) hardware components \cite{mMIMO_survey}.
In order to achieve the benefits of massive MIMO, the accurate channel estimation technique is of vital.
However, channel estimation is challenging in massive MIMO systems since pilot length for downlink channel estimation in frequency division duplex (FDD) is proportional to the number of antennas as the BS.
In contrast, the pilot overhead which is proportional to the number of user equipments (UEs) can be significantly reduced by exploiting the channel reciprocity in time-division duplex (TDD) mode \cite{JSTSP_MIMO}.
Despite the use of TDD mode, the pilot overhead in multi-cell scenario has to be proportional to the number of all UEs in all the cells for allocating orthogonal pilots to UEs \cite{ACSSC_num_pilot}. In practical systems, however, the allocated pilot sequences are no longer orthogonal between UEs in adjacent cells since the pilot length needs to be limited by coherence time. As a result, it leads to pilot contamination which is a fundamentally limiting factor degrading the channel estimation performance \cite{TCOM_toward,TWC_pilot_contamin,TWC_pilot_contamin2}.
Furthermore, the pilot contamination induces inter-cell interference due to the reuse of pilot sequences in multi-cell environments, which deteriorates the system throughput \cite{ISIT_effect_pilot}.

Due to the feature of BS with a few hundreds of antennas, BS can simultaneously serve a large number of UEs by exploiting the spatial multiplexing.
To implement massive antenna components at BS, it is attractive to mount a low-cost and power-efficient antenna component because hardware cost and power consumption grow linearly as a function of the number of antennas.
However, the use of a cheap component is likely to introduce hardware impairments such as non-linear power amplifier, phase noise, and I/Q imbalance \cite{T_RHWI,RHWI,RHWI2}.
Unfortunately, the effects of hardware impairments cannot be perfectly removed and a certain amount of residual hardware impairments (RHWIs) always remains even though the compensation or calibration techniques are utilized \cite{T_RHWI}.
As a result, the distortion noise caused by RHWIs is also the source of pilot contamination as well as the lack of orthogonal pilot sequences, which degrades the channel estimation performance \cite{mMIMO_survey}.
In order to minimize the effects of pilot contamination, we focus on the pilot sequence design and the distortion noise mitigation caused by pilot contamination to obtain the accurate channel state information (CSI).
 
\subsection{Related Work}

Several strategies for mitigating pilot contamination have been studied in terms of pilot design.
Different from the orthogonal based pilot design, the non-orthogonal based pilot design significantly improves the channel estimation performance \cite{random_pilot}.
Non-orthogonal pilot design has been investigated in \cite{NO_pilot4,NO_pilot5,NO_pilot1,NO_pilot3} to minimize mean square error (MSE) with linear minimum mean square error (LMMSE) channel estimator but the minimizing MSE is non-convex problem.
To handle the non-convexity of MSE, the non-convex problem is decomposed into distributed sub-problems and successive optimization approach is adopted to solve the sub-problems \cite{NO_pilot4}.
Moreover, authors in \cite{NO_pilot5} suggested greedy algorithm 
to obtain the pilot sequences and a linearized alternating direction method of multipliers (L-ADMM) algorithm and fractional programming (FP) were introduced in \cite{NO_pilot1} and \cite{NO_pilot3}, respectively, where these methods are used to approximate the MSE function in a linear form.
However, all of these works do not consider hardware impairments and it cannot guarantee the optimal solution due to the approximation of non-convex problem.
On the other attempts, deep learning based pilot design has recently been introduced in \cite{WCL_orth_pilot, TNN,TWC_prun}.
Authors in \cite{WCL_orth_pilot} considered the power allocation of pilot sequences to minimize MSE using deep learning approach.
However, non-orthogonal pilot scheme offers better performance than \cite{WCL_orth_pilot} since it restricts the pilot sequences to be orthogonal.
In \cite{TNN, TWC_prun}, deep learning based joint pilot design and channel estimation was proposed but it does not address the pilot contamination in multi-cell environment.
Therefore, the pilot design is crucial to obtain the accurate CSI and we develop a deep learning based pilot design to suppress the effects of pilot contamination.

Another performance bottleneck of massive MIMO is hardware impairments which worsen the effects of pilot contamination.
Although hardware impairments can be compensated by using analog and digital signal processing \cite{HI_compen}, there always exists a certain amount of RHWIs that causes the distortion noise and degrades the channel estimation performance \cite{T_RHWI, RHWI}.
To address the problem of pilot contamination caused by non-orthogonal pilot and RHWIs for channel estimation, LMMSE channel estimator is widely used by balancing between interference suppression and noise enhancement \cite{RHWI_main}.
Moreover, deep learning based channel estimation with pilot contamination has recently investigated in \cite{OJVT_contamination, OJCS_HI}.
However, these works as well as LMMSE estimator require the prior knowledge of channel statistics and RHWIs. 
In practical systems, the distortion noise caused by RHWIs is unintended and inevitable and the prior knowledge about RHWIs cannot be obtained.
Therefore, channel estimation has to be done without the prior knowledge of RHIWs, which requires a technique to eliminate the unknown effects of pilot contamination caused by RHWIs.

It is well-known that deep residual learning can estimate the unknown noise which subtracts from the noisy observations to construct a original data \cite{Conventional_Filter1}.
Deep residual learning has been also investigated for wireless communications in \cite{denoiser1, denoiser2}.
Authors in \cite{denoiser1,denoiser2} model the channel estimation problem as a denoising problem and develop convolutional neural network (CNN) based deep residual learning for denoiser to recover the channel from the noisy received pilot signal. In this model, denoising block supports the channel estimation by learning the residual noise from the received signal.
However, they only consider additive white Gaussian noise (AWGN) but the distortion noise caused by RHWIs is non-Gaussian noise.
According to \cite{RealNoise_Filter1}, denoising CNN (DnCNN) cannot effectively suppress the real noisy data. In order to solve this problem,  adaptive instance normalization denoising network (AINDNet) was proposed for image processing using transfer learning \cite{CVPR_2020_AINDNet}. It has the ability to denoise the non-Gaussian noise by adopting adaptive instance normalization residual block (AIN-ResBlock).
Motivated by this, we adopt the AIN-ResBlock model for deep residual learning to denoise the non-Gaussian noise caused by RHWIs. 

\subsection{Contributions}
To handle the effects of pilot contamination caused by non-orthogonal pilot sequences and RHWIs, the statistics of RHWIs and channel between the neighboring cells need to be known for channel estimator.
However, it is difficult to acquire the statistics of them, and thus channel estimation should be performed without the knowledge of pilot contamination.
Therefore, we propose the joint pilot design and channel estimation using deep residual learning to mitigate the effects of pilot contamination when the prior knowledge cannot be exploited.
To best our knowledge, the joint pilot design and channel estimator for multi-cell massive MIMO systems with hardware impairments has not been considered.
The major remarkable contributions of this paper can be summarized as follows.
\begin{itemize}
    \item We first propose a deep learning based pilot design to minimize MSE of LMMSE estimator. Using the derived MSE result, our proposed neural network is trained by unsupervised learning that does not require the channel samples for training. The neural network based pilot design model is also implemented to the joint pilot design and channel estimator to cope with the effects of pilot contamination. 
    \item In addition, we adopt a deep residual learning for channel estimator to alleviate the distortion noise caused by pilot contamination.
    By learning the distortion noise from the received signal without the prior knowledge, proposed estimator can effectively construct the original channel. We confirm that proposed estimator achieves almost the same performance with LMMSE estimator and it particularly outperforms LMMSE with the existence of RHWIs.
    
    \item Lastly, we propose a novel joint pilot design and channel estimator by adopting the transfer learning. The pilot design model is firstly trained for LMMSE criterion using unsupervised learning approach without the prior knowledge of pilot contamination.
    Using the pre-trained model, pilot design and channel estimator are jointly trained to minimize MSE of estimated channel. 
    As a result, it outperforms the conventional LMMSE estimator with orthogonal pilot reuse scheme even though the proposed scheme does not exploit the prior knowledge.
\end{itemize}

\subsection{Organization}
The rests of this paper are organized as follows.
Sec.~\ref{sec:model} shows signal and system models.
Sec.~\ref{sec:pilot_design} presents the design of non-orthogonal pilot sequences and Sec.~\ref{sec:channel_estimator} proposes channel estimation using deep residual learning.
Based on these results, joint pilot and channel estimator design is presented in Sec.~\ref{sec:joint}.
Sec.~\ref{sec:simul} shows the performance evaluation results and Sec.~\ref{sec:conclusion} concludes this paper.

\section{Signal and System Models}\label{sec:model}
\subsection{System Model}
Consider a multi-cell massive MIMO system with $L$-cells where
each BS is equipped with $N$ antennas and $K$ single-antenna UEs are deployed in each cell.
Assuming TDD mode, the uplink and downlink channels can be estimated using uplink pilot sequences due to channel reciprocity.
In addition, block fading is assumed where channels remain static within coherence time, $T$, and the channel varies independently during different coherence time intervals.
\begin{figure}[t]\centering
\includegraphics[width=80mm]{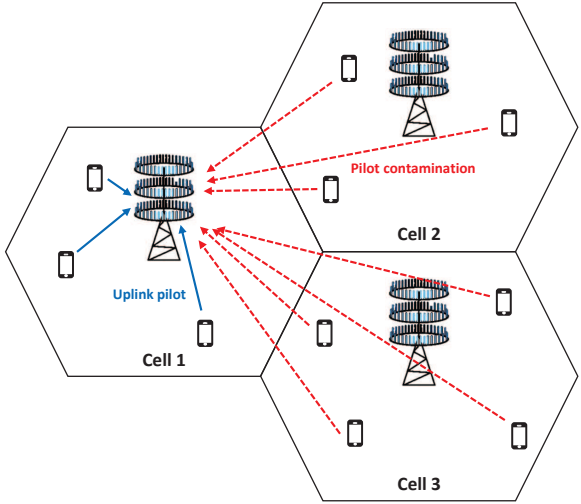}
\caption{Illustration of multi-cell massive MIMO systems with pilot contamination}\label{fig1}
\end{figure}

As in Fig. \ref{fig1}, pilot contamination is caused by non-orthogonal pilot sequences across multi-cells during the uplink training phase. In addition, pilot sequence is also contaminated by hardware impairments which are inevitable in practical wireless systems \cite{mMIMO_survey}. 
For massive MIMO systems, 
low-cost radio frequency (RF) hardware component is desirable which makes the the effects of hardware impairments on the performance worse.
Therefore, hardware impairments due to the low-cost device should be considered, which causes a mismatch with the desired signals and the transmitted signals over the channel.
Although hardware impairments can be mitigated with the advanced calibration or compensation algorithms, there still exists the RHWIs which deteriorate the link performance.
In the existence of RHWIs, the received signal at $\mathrm{BS}_i$ in the $i$-th cell for uplink training in order to estimate CSI can be written as
\begin{align}\label{y_HI}\notag
    \mathbf{Y}_i=&\sum\limits_{j=1}^L\sum\limits_{k=1}^K \mathbf{h}_{i,j,k}\left(\sqrt{\tau_p P_{j,k}}\mathbf{x}_{j,k}^H+\boldsymbol{\eta}_{\mathrm{UE}_{j,k}}^H \right)+\boldsymbol{\eta}_{\mathrm{BS}_i}+\mathbf{n}_i\\\notag
    =&\sqrt{\tau_p}\mathbf{H}_{i,i}\left(\mathbf{P}_i^{\frac{1}{2}}\mathbf{X}_i^H+\boldsymbol{\eta}_{\mathrm{UE}_i}^H\right)\\
    &+\sqrt{\tau_p}\sum\limits_{\substack{j=1\\j\neq i}}^L\mathbf{H}_{i,j}\left(\mathbf{P}_j^{\frac{1}{2}}\mathbf{X}_j^H+\boldsymbol{\eta}_{\mathrm{UE}_j}^H\right)+\boldsymbol{\eta}_{\mathrm{BS}_i}+\mathbf{n}_i,
\end{align}
where $\mathbf{H}_{i,j}=\left[\mathbf{h}_{i,j,1},\cdots,\mathbf{h}_{i,j,K}\right]$ is the channel matrix between all users in the $j$-th cell and the $\mathrm{BS}_i$ and each column $\mathbf{h}_{i,j,k}=\sqrt{\beta_{i,j,k}}\mathbf{g}_{i,j,k}$ denotes the channel response vector from the $k$-th UE in the $j$-th cell, denoted as  $\mathrm{UE}_{j,k}$, to the $\mathrm{BS}_i$. In the channel response vector, $\beta_{i,j,k}$ denotes the large-scale fading including both path loss and shadowing.\footnote{Since the large scale fading changes slower than small-scale fading, we assume that $\beta_{i,i,k}$ is known prior at the $\mathrm{BS}_i$ and it is available during the channel estimation phase.}
In addition, $\mathbf{g}_{i,j,k}\in \mathbb{C}^{N\times 1}$ denotes the small-scale fading of which elements are modeled as independent and identical distributed (i.i.d.) Rayleigh fading, i.e., $\mathbf{g}_{i,j,k} \sim \mathcal{CN}(\mathbf{0},\mathbf{I}_{N})$.
Moreover, $\mathbf{P}_i=\mathrm{diag}\left(\left[P_{i,1},\cdots,P_{i,K}\right]\right)$ is the uplink transmit power matrix where $P_{i,k}$ is the uplink transmit power of $\mathrm{UE}_{i,k}$, $\mathbf{X}_i=\left[\mathbf{x}_{i,1},\cdots,\mathbf{x}_{i,K}\right]$ is the collection of pilot sequences where $\mathbf{x}_{i,k}\in \mathbb{C}^{\tau_p\times 1}$ indicates the pilot sequence with the length of $\tau_p$ with $\mathbf{x}_{i,k}^H\mathbf{x}_{i,k}=1$, and $\mathbf{n}_i\in \mathbb{C}^{N\times \tau_p}$ is the AWGN where each column is distributed as $\mathcal{CN}(\mathbf{0},\sigma^2\mathbf{I})$.
It is worthy noting that the condition of pilot sequences is not restricted to $\mathbf{x}_{i,k}^H\mathbf{x}_{i,k'}=0,~\forall~k\neq k'$ for the non-orthogonal pilot design.

As shown in \eqref{y_HI}, the distortion noise incurred by RHWI at UE and BS in the $i$-th cell are denoted as $\boldsymbol{\eta}_{\mathrm{UE}_i}\in \mathbb{C}^{\tau_p \times K}$ and $\boldsymbol{\eta}_{\mathrm{BS}_i}\in \mathbb{C}^{N\times \tau_p}$, respectively.
It is revealed that the distortion noise can be modeled as a Gaussian noise by aggregating the effects of RHWIs where the variance of distortion noise is proportional to the signal powers and the channel gain~\cite{RHWI}.
We assume that the distortion noise is independent among different antennas, UEs, and time~\cite{T_RHWI}. Thus, the distortion noise induced by RHWIs at $\mathrm{UE}_{i,k}$ and $\mathrm{BS}_i$ can be modeled as
\begin{align}
    &\boldsymbol{\eta}_{\mathrm{UE}_{i,k}} \sim \mathcal{CN}\left(0,\delta_{\mathrm{UE}_{i,k}}^2P_{i,k}\mathbf{I}_{\tau_p}\right),\\
    &\boldsymbol{\eta}_{\mathrm{BS}_i,t} \sim \mathcal{CN}\left(0,\delta_{\mathrm{BS}_i}^2\sum\limits_{j=1}^L\mathrm{diag}\left(\mathbf{H}_{i,j}\mathbf{P}_j\mathbf{H}_{i,j}^H\right)\right),
\end{align}
where $\boldsymbol{\eta}_{\mathrm{BS}_i,t}$ is the $t$-th column of $\boldsymbol{\eta}_{\mathrm{BS}_i}$ with  $\mathbb{E}\left[\boldsymbol{\eta}_{\mathrm{BS}_i,t}\boldsymbol{\eta}_{\mathrm{BS}_i,t'}^H\right]=\mathbf{0},~ \forall t\neq t'$,
and $\delta_{\mathrm{UE}_{i,k}}$ and $\delta_{\mathrm{BS}_i}$ denote the proportional coefficients indicating the level of RHWIs in the transmitter and receiver, respectively.
We note that the level of RHWIs can be measured by a certain metric such as the error vector magnitude (EVM) in a practical system.
For example, the EVM at $\mathrm{UE}_{i,k}$ is defined as follows~\cite{T_RHWI}:
\begin{align}
    \mathrm{EVM}_{\mathrm{UE},i,k}=\sqrt{\frac{\mathbb{E}\left[||\boldsymbol{\eta}_{\mathrm{UE}_{i,k}}||^2\right]}{\mathbb{E}\left[||\sqrt{\tau_pP_{i,k}}\mathbf{x}_{i,k}||^2\right]}}=\delta_{\mathrm{UE}_{i,k}}.
\end{align}
Typically, EVM is often adopted to measure the quality of transceivers, and 3GPP LTE specifies the EVM requirements in the range $[0.08, 0.175]$~\cite{LTE_EVM}.
However, the larger EVMs are of the interest in a massive MIMO system because the relatively low-cost antenna components are taken into account. We here consider the level of RHWIs in the range $[0, 0.2]$.

\subsection{Conventional LMMSE Channel Estimator}

Each BS should acquire the CSI of individual UEs to perform the advanced receiver techniques. In the channel estimation phase, each UE transmits its own pilot sequence at the same time and BS corresponding to each UE estimates the channels using the known pilot sequences from each UE. 
In general, the pilot orthogonality is guaranteed with $K<\tau_p$ whereas the orthogonal pilot sequences need to be shared with the neighboring cells due to the limited coherence time, which cannot ensure the orthogonality among all the cells. Hence, the non-orthogonal pilot design is essential to mitigate the effects of pilot contamination in multi-cell scenario.
When the non-orthogonal pilot sequences are used, the received signal in \eqref{y_HI} can be rewritten as
\begin{align}\label{rx_signal}
    \mathbf{Y}_i=\sqrt{\tau_p}\mathbf{H}_{i,i}\mathbf{P}_i^{\frac{1}{2}}\mathbf{X}_i^H+\underbrace{\sqrt{\tau_p}\sum\limits_{\substack{j=1\\j\neq i}}^L\mathbf{H}_{i,j}\mathbf{P}_j^{\frac{1}{2}}\mathbf{X}_j^H}_{\mathrm{Inter-cell ~interference}}+\tilde{\mathbf{N}}_i,
\end{align}
where $\tilde{\mathbf{N}}_i=\sqrt{\tau_p}\sum_{j=1}^L\mathbf{H}_{i,j}\boldsymbol{\eta}_{\mathrm{UE}_j}^H+\boldsymbol{\eta}_{\mathrm{BS}_i}+\mathbf{n}_i$ is the distortion noises caused by RHWIs at UEs and BS.
Note that the second term in \eqref{rx_signal} is inter-cell interference caused by pilot contamination. It can be eliminated using the orthogonal pilot sequences in the different cells but the sufficient pilot duration should be guaranteed.
For given channel realization, the covariance matrix of $\tilde{\mathbf{N}}_i$ can be calculated as
\begin{align}\notag\label{cov_distort}
    \mathbf{K}_{\tilde{N}_i}=&\mathbb{E}[\tilde{\mathbf{N}}_i\tilde{\mathbf{N}}_i^H]\\\notag
    =&\tau_p^2\sum_{j=1}^L\mathbf{H}_{i,j}\mathbf{\Delta}_{\mathrm{UE}_j}\mathbf{P}_{j}\mathbf{H}_{i,j}^H\\
    &+\tau_p\delta_{\mathrm{BS}_i}^2\sum\limits_{j=1}^L\mathrm{diag}\left(\mathbf{H}_{i,j}\mathbf{P}_j\mathbf{H}_{i,j}^H\right)+\tau_p\sigma^2\mathbf{I},
\end{align}
where $\mathbf{\Delta}_{\mathrm{UE}_i}=\mathrm{diag}\left([\delta_{\mathrm{UE}_{i,1}}^2,\cdots,\delta_{\mathrm{UE}_{i,K}}^2]\right)$.
Due to the RHWI at UEs, the aggregated distortion noise is represented as colored noise with covariance matrix, $\mathbf{K}_{\tilde{N}_i}$.

To estimate the channel using \eqref{rx_signal}, least square (LS) estimator can be applied such as $\frac{1}{\sqrt{\tau_p}}\mathbf{Y}_i\mathbf{X}_i(\mathbf{X}_i^H\mathbf{X}_i)^{-1}$. It is simple to implement since it does not require the prior knowledge of channel statistics.
However, the performance of LS estimator is worse than that of Bayesian estimator such as MMSE estimator.
LMMSE estimator is widely used for channel estimation which achieves the optimal performance in terms of minimizing MSE if $\mathbf{Y}_i$ and $\mathbf{H}_{i,i}$ are jointly Gaussian.
However, $\mathbf{H}_{i,j}\boldsymbol{\eta}_{\mathrm{UE}_j}^H$ term in \eqref{rx_signal} has double Gaussian distribution which makes LMMSE estimator not optimal. On the other hand, it is difficult to obtain the general MMSE estimator, $\hat{\mathbf{H}}=\mathbb{E}[\hat{\mathbf{H}}|\mathbf{Y}]$, as a closed-form \cite{MMSE_book}.
As an alternative approach, we adopt LMMSE estimator which can be derived in a closed form, which provides better performance than LS estimator. 

At the uplink training phase, the channel can be estimated using LMMSE estimator as
\begin{align}\label{LMMSE}\notag
    \hat{\mathbf{H}}_{i,i}=&\mathbf{Y}_i\mathbf{A}_i\\
    =&\frac{1}{\sqrt{\tau_p}}\mathbf{Y}_i\left(\sum\limits_{j=1}^L\mathbf{X}_j\mathbf{D}_{i,j}\mathbf{P}_j\mathbf{X}_j^H+\phi_i\mathbf{I}\right)^{-1}\mathbf{X}_i\mathbf{P}_i^{\frac{1}{2}}\mathbf{D}_{i,i},
\end{align}
where $\phi_i=(\sum_{j=1}^L\sum_{k=1}^K(\tau_p\delta^2_{\mathrm{UE},j,k}+\delta^2_{\mathrm{BS},i})\beta_{i,j,k}P_{j,k}+\sigma^2)/\tau_p$, $\mathbf{A}_i$ is the LMMSE estimator for the $i$-th cell, $\mathbf{G}_{i,j}=\left[\mathbf{g}_{i,j,1},\cdots,\mathbf{g}_{i,j,K}\right]$, and $\mathbf{D}_{i,j}=\mathrm{diag}\left([\mathbf{\beta}_{i,j,1},\cdots,\mathbf{\beta}_{i,j,K}]\right)$ (i.e., $\mathbf{H}_{i,j}=\mathbf{G}_{i,j}\mathbf{D}_{i,j}^{1/2}$).
In addition, the MSE using LMMSE estimator can be represented as
\begin{align}
    \mathrm{MSE}_{i}=N\mathrm{Tr}\left(\left(\mathbf{D}_{i,i}^{-1}+\mathbf{X}_i^H\mathbf{B}_i^{-1}\mathbf{X}_i\mathbf{P}_i\right)^{-1}\right),
\end{align}
where $\mathbf{B}_i=\sum_{j=1, j\neq i}^L\mathbf{X}_j\mathbf{D}_{i,j}\mathbf{P}_j\mathbf{X}_j^H+\phi_i\mathbf{I}$.

\textit{Proof: See Appendix A}

It is worth noting that \eqref{LMMSE} can be obtained using the prior knowledge of channel statistics and RHWIs.
However, it is infeasible to acquire the perfect knowledge of inter-cell large-scale fading coefficients because it requires the estimation of $(L-1)K$ coefficients at each BS which has prohibitively high overhead in massive connectivity \cite{IC_pathloss}.
Furthermore, the exact level of RHWIs is not available due to the imperfect compensation of hardware impairments.
Considering all those practical limitations,
we propose the pilot design and deep residual learning based channel estimator that does not require the prior knowledge.

\section{Design of Non-Orthogonal Pilot Sequences}\label{sec:pilot_design}

In general, the orthogonal pilot sequences are desirable to eliminate the inter-cell interference. However, 
the orthogonal pilot sequences may not be available for the case of massive connectivity such as \cite{ComMag_IoT} due to the relatively short coherence time such as $K>\tau_p$.
Even if the orthogonal pilot sequences are used in a cell, there is pilot contamination issue due to the shortage of orthogonal pilot sequences between all the cells.
Thus, it is better to design the non-orthogonal pilot sequences than the reuse of orthogonal pilot sequences in the different cells to support multi-cell scenario.
Therefore, we need to design the non-orthogonal pilot sequences to resolve the shortage of orthogonal pilot sequences as well as to mitigate the effects of pilot contamination.

Based on the derived LMMSE estimator, we need to solve the pilot design problem given as
\begin{align}\notag\label{P1_design}
    \underset{\mathbf{X}_i,\mathbf{P}_i}{\min} &\sum\limits_{i=1}^L N\mathrm{Tr}\left(\left(\mathbf{D}_{i,i}^{-1}+\mathbf{X}_i^H\mathbf{B}_i^{-1}\mathbf{X}_i\mathbf{P}_i\right)^{-1}\right)\\
    \mathrm{s.t}~&\mathbf{x}_{i,k}^H\mathbf{x}_{i,k}=1,~\forall~1\leq i\leq L, 1\leq k\leq K\\\notag
    &P_{i,k}\leq P_\mathrm{max}
\end{align}
where $P_\mathrm{max}$ is the maximum transmit power of UE.
By setting $\Bar{\mathbf{X}}_i=\mathbf{X}_i\mathbf{P}_i^{1/2}$, \eqref{P1_design} can be reformulated as
\begin{align}\notag\label{P2_design}
    \underset{\bar{\mathbf{X}}_i}{\min} &\sum\limits_{i=1}^L N\mathrm{Tr}\left(\left(\mathbf{D}_{i,i}^{-1}+\bar{\mathbf{X}}_i^H\bar{\mathbf{B}}_i^{-1}\bar{\mathbf{X}}_i\right)^{-1}\right)\\
    \mathrm{s.t}~&\bar{\mathbf{x}}_{i,k}^H\bar{\mathbf{x}}_{i,k}\leq P_{max},~\forall~1\leq i\leq L, 1\leq k\leq K
\end{align}
where $\bar{\mathbf{B}}_i=\sum_{j=1, j\neq i}^L\bar{\mathbf{X}}_j\mathbf{D}_{i,j}\bar{\mathbf{X}}_j^H+\phi_i\mathbf{I}$.
Note that \eqref{P2_design} is a non-convex problem and it is difficult to obtain the global optimum. As in \cite{NO_pilot1,NO_pilot3,NO_pilot4}, there are some methods to solve \eqref{P2_design} but unfortunately they cannot guarantee the global optimum by approximating the objective function.
Also, it is hard to implement in practical systems due to the varying level of RHWIs and their unknown parameter values.
Therefore, we propose the deep learning based non-orthogonal pilot design for LMMSE estimator to minimize MSE.

\subsection{Deep Neural Network Based Pilot Design Structure}
\begin{figure*}[t]\centering
\includegraphics[width=170mm]{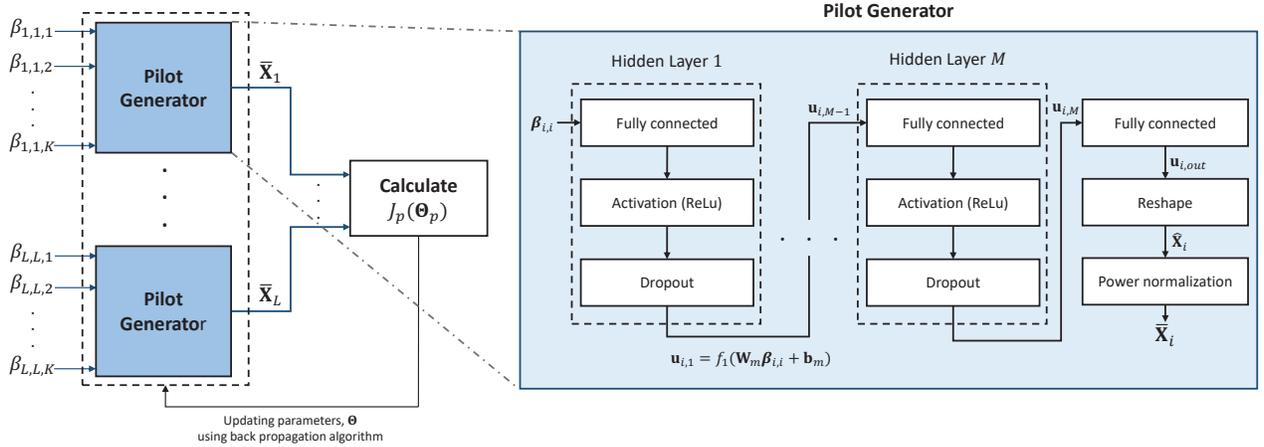}
\caption{The structure of proposed non-orthogonal pilot design for LMMSE estimator}\label{fig2}
\end{figure*}
As in \eqref{P2_design}, the optimal pilot sequences are different when the large-scale fading of UEs is changed. Thus, each cell needs to generate the pilot sequences using the large-scale fading information of UEs. However, it is difficult to acquire the information of inter-cell and other cells large-scale fading coefficients in each cell. Hence, our aim is to design the neural network to independently generate the pilot sequences using only local large-scale fading coefficients in each cell.

To address the design of non-orthogonal pilot sequences minimizing MSE as in \eqref{P2_design}, we adopt the fully connected neural network structure as in Fig. \ref{fig2}. 
The structure of proposed neural network based pilot design is composed of the input layer, the $M$ hidden layers, and the output layer.
For the input of pilot generator, its own large-scale fading coefficients are used to obtain the pilot sequences independently when it is implemented in on-line.
In the $m$-th hidden layer with $n_m$ nodes, the output of hidden layer can be expressed as
\begin{align}
    \mathbf{u}_{i,m}=f_m(\mathbf{W}_m\mathbf{u}_{i,m-1}+\mathbf{b}_m),
\end{align}
where $\mathbf{W}_m\in \mathbb{R}^{n_m\times n_{m-1}}$ and $\mathbf{b}_m\in \mathbb{R}^{n_m\times 1}$ are the weight matrix and the bias vector at the $m$-th hidden layer, respectively, and $f_m(\cdot)$ denotes the function including activation and dropout in the $m$-th hidden layer.
Without loss of generality, $\mathbf{u}_{i,0}$ denotes the input of pilot generator, $\boldsymbol{\beta}_{i,i}=[\beta_{i,i,1},\cdots,\beta_{i,i,K}]$.

After passing through the hidden layers, the output dimension should be matched to the dimension of pilot sequences using the fully connected output layer which can be written as 
\begin{align}
    \mathbf{u}_{i,\mathrm{out}}=\mathbf{W}_\mathrm{out}\mathbf{u}_{i,M}+\mathbf{b}_\mathrm{out},
\end{align}
where $\mathbf{W}_\mathrm{out}\in \mathbb{R}^{2\tau_pK\times n_{M}}$ and $\mathbf{b}_\mathrm{out}\in \mathbb{R}^{2\tau_pK\times 1}$ are the output weights and bias, respectively.
Lastly, real-valued output, $\mathbf{u}_{i,\mathrm{out}}$, is reshaped into complex-valued pilot matrix $\tilde{\mathbf{X}}_i\in \mathbb{C}^{\tau_p \times K}$.
However, the output pilot sequence, $\tilde{\mathbf{X}}_i$, does not satisfy the power constraint in \eqref{P2_design}. Thus, we also employ a normalization layer following the fully connected output layer to satisfy the power constraint. As in \cite{TNN}, projected gradient descent (PGD) method can be applied to meet the power constraint when the pilot sequences are mapped into the weights of neural network. Since the weights of neural network are not pilot sequences in our proposed model unlike \cite{TNN}, we utilize a power normalization function in the proposed model alternatively.
The power normalization operates as follows,
\begin{align}
    f_\mathrm{norm}(\tilde{\mathbf{X}}_i)=\tilde{\mathbf{X}}_i\tilde{\mathbf{P}}_i,
\end{align}
where $\tilde{\mathbf{P}}_i$ is the diagonal matrix and its $k$-th element is given as
\begin{align}
    [\tilde{\mathbf{P}}_i]_{k,k}=\left\{\begin{matrix*}[l]
    1, & \mathrm{if}~ \|\tilde{\mathbf{x}}_{i,k}\|^2\leq P_k \\
    \sqrt{P_k}/\|\tilde{\mathbf{x}}_{i,k}\|, & \mathrm{if}~ \|\tilde{\mathbf{x}}_{i,k}\|^2>P_k
    \end{matrix*}\right.
\end{align}
where $\tilde{\mathbf{x}}_{i,k}$ is the $k$-th column of $\tilde{\mathbf{X}}_i$ As a result, the pilot sequences for the $i$-th cell can be designed as $\hat{\mathbf{X}}_{i}=f_\mathrm{norm}(\tilde{\mathbf{X}}_i)$.


\subsection{Training of Pilot Design}
As the input of proposed pilot design model, the local large-scale fading coefficients with the length of $K$ are used. The real-valued output of proposed model with the dimension of $2\tau_pK$ is reshaped into complex-valued matrix and its power is normalized to construct the desired pilot sequences, $\hat{\mathbf{X}}_i$.
In the off-line training phase, we train the parameters of pilot design model, $\boldsymbol{\Theta}_p$, to minimize the loss function as follows,
\begin{align}\label{p_loss}
    J_p(\boldsymbol{\Theta}_p)=\frac{1}{LN_T}\sum\limits_{n=1}^{N_T}\sum\limits_{i=1}^L \mathrm{Tr}\left(\left(\mathbf{D}_{i,i}^{{(p)}^{-1}}+\hat{\mathbf{X}}_i^{(p)^H}\bar{\mathbf{B}}_i^{(p)^{-1}}\hat{\mathbf{X}}_i^{(p)}\right)^{-1}\right),
\end{align}
where $\boldsymbol{\Theta}_p=\{\mathbf{W}_m, \mathbf{b}_m,\mathbf{W}_\mathrm{out}, \mathbf{b}_\mathrm{out} \}_{m=1}^M$ denotes the set of parameters and $N_T$ is the number of training samples.
Note that the loss function in \eqref{p_loss} does not require the instantaneous true channel but only statistics of channel since the loss function is obtained by averaging over the channel realization.
Hence, there is no need to acquire the full channel information for training and it has the advantage of training pilot design model by using the statistics of channel.

After the phase of training pilot design model, the weights of trained model are distributed to each cell for on-line pilot generation. Each cell can generate its own pilot sequences independently using the local large-scale fading even though the inter-cell large-scale fading cannot be exploited in on-line phase. 
Although the proposed pilot design model does not require the true channel information, all the large-scale fading coefficients and the level of RHWIs need to be known to calculate the loss function.
Thus, it cannot be directly applied to real scenario due to the practical limitation of obtaining the prior knowledge.
To address the issue of practical implementation, we will discuss pilot contamination unaware pilot design model in Sec.~\ref{sec:joint}.


\section{Deep Residual Learning Based Channel Estimator}\label{sec:channel_estimator}
It is obvious that interference caused by pilot contamination deteriorates the channel estimation performance as in \eqref{rx_signal}. 
It should be noted that LMMSE estimator is designed using the statistics of channel (i.e., large-scale fading) and the level of RHWIs. 
However, it is challenging to obtain the prior knowledge for LMMSE estimator.
Moreover, LMMSE estimator cannot provide the optimal performance in the existence of RHWIs, although the prior knowledge can be exploited for the channel estimator.
Therefore, it is infeasible to apply LMMSE estimator in practical systems.
 As an alternative approach, we propose the deep residual learning based channel estimator to alleviate the effects of pilot contamination caused by non-orthogonal pilot sequences and RHWIs.
In \cite{denoiser2}, deep residual learning is exploited to estimate the channel by modeling as a denoising problem.
On the other hand, the distortion noise caused by pilot contamination is a complicated model rather than a simple Gaussian model.
Thus, we need to adopt another deep residual learning approach to cope with the non-Gaussian noise model.
In \cite{CVPR_2020_AINDNet}, a generalized denoising architecture based on AIN-ResBlock is proposed, which can learn the general features of noise.
Thus, we adopt the structure of AIN-ResNet and it can be utilized for eliminating the distortion noise caused by pilot contamination.


\subsection{Pre-processing}
Instead of LMMSE estimator, we utilize LS estimator that does not require the prior knowledge of channel statistics and RHWIs.
Note that LS estimator can be applied to the case of $\tau_p\geq K$. Then, LS estimator can be given by ${\mathbf{A}}_i^\mathrm{LS}=\frac{1}{\sqrt{\tau_p}}\bar{\mathbf{X}}_i(\bar{\mathbf{X}}_i^H\bar{\mathbf{X}}_i)^{-1}$ and the estimated channel can be written as
\begin{align}\notag\label{denoise_model}
    \hat{\mathbf{Y}}_{i,i}&=\mathbf{Y}_i{\mathbf{A}}_i^\mathrm{LS}\\\notag
    &=\mathbf{H}_{i,i}+\underbrace{\sum\limits_{\substack{j=1\\j\neq i}}^L\mathbf{H}_{i,j}\bar{\mathbf{X}}_j^H{\mathbf{A}}_i^\mathrm{LS}+\tilde{\mathbf{N}}_i{\mathbf{A}}_i^\mathrm{LS}}_\mathrm{Distortion~noise}\\
    &=\mathbf{H}_{i,i}+\mathbf{Z}_i,
\end{align}
where $\mathbf{Z}_i={\sum_{j=1, j\neq i}^L\mathbf{H}_{i,j}\bar{\mathbf{X}}_j^H\mathbf{A}_i^\mathrm{LS}+\tilde{\mathbf{N}}_i \mathbf{A}_i^\mathrm{LS}}$ denotes the distortion noise induced by pilot contamination.
It is note that \eqref{denoise_model} can be regraded as a denoising problem since $\hat{\mathbf{Y}}_{i,i}$ consists of the desired channel and noise terms.
Therefore, our aim is to denoise the distortion noise term, $\mathbf{Z}_i$, using deep residual learning to recover the desired channel.
\subsection{Architecture of Channel Estimator}
\begin{figure*}[t]\centering
\includegraphics[width=170mm]{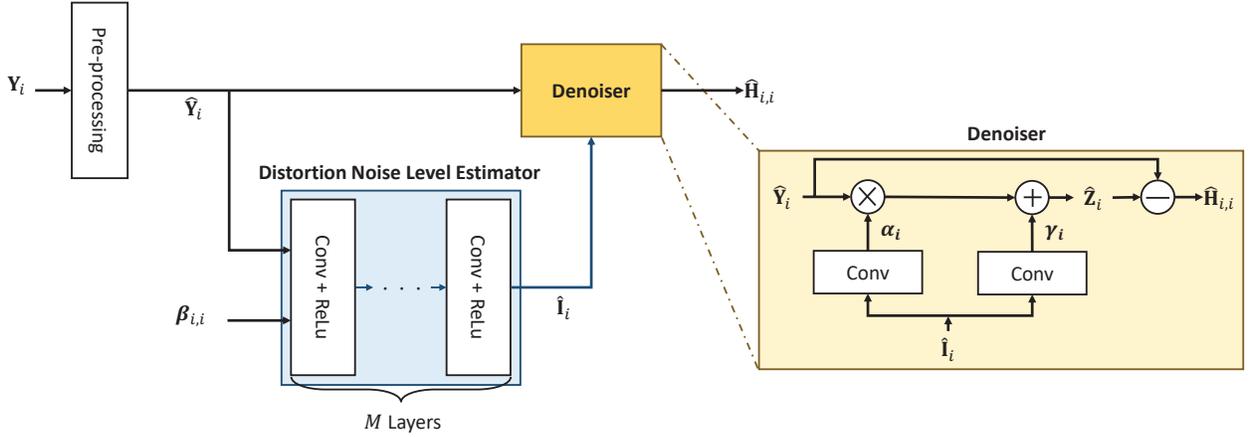}
\caption{The structure of deep residual learning based channel estimator}\label{fig3}
\end{figure*}

To suppress the distortion noise, we adopt deep residual learning suggested in \cite{CVPR_2020_AINDNet} where the neural network consists of two components: noise estimator and denoiser. Although it was suggested for image processing, it can be applied for channel estimator since channel can be treated as an image. To implement the structure of deep residual learning to multi-cell massive MIMO systems, we adjust the neural network architecture by reducing unnecessary modules and blocks. The whole structure of proposed channel estimator is shown in Fig.~\ref{fig3} where the proposed channel estimator consists of the distortion noise level estimator and denoiser. 

For the input of deep residual learning, the pre-processed signal, $\hat{\mathbf{Y}}_{i,i}$, is first separated with the real and imaginary parts (i.e., $\textrm{Re}(\hat{\mathbf{Y}}_{i,i})$ and $\textrm{Im}(\hat{\mathbf{Y}}_{i,i})$).
In addition, we add the local large-scale fading coefficients, $\boldsymbol{\beta}_{i,i}$, as an input to extract the features from the pre-processed signal. 
Note that $\boldsymbol{\beta}_{i,i}$ with the length of $K$ needs to be reshaped into $N\times K$ matrix to employ as an input.
Typically, the large-scale fading coefficient is the same over different antennas at a BS. Then, the local large-scal fading can be expanded as $\mathbf{1}_N\boldsymbol{\beta}_{i,i}^T$ where $\mathbf{1}_N$ represents the all in one column vector with the size of $N$.
Thus, we can construct an input as $\tilde{\mathbf{Y}}_{i,i}=\{\mathrm{Re}(\hat{\mathbf{Y}}_{i,i}), \mathrm{Im}(\hat{\mathbf{Y}}_{i,i}),\mathbf{1}_N\boldsymbol{\beta}_{i,i}^T\}\in \mathbb{R}^{N\times K \times 3}$. 
As in Fig. \ref{fig3}, the distortion noise level estimator is composed of $M$ convolutional layers with rectified linear unit (ReLu) activation function, which is used to extract the features of the distortion noise. 
For this estimator, we use 64 filters with the size of $3\times 3\times 3$ for the first layer to meet the input size and  64 filters with the size of $3\times 3\times 64$, otherwise.
In denoiser block illustrated in Fig. \ref{fig3}, the scaling factor, $\boldsymbol{\alpha}_i\in \mathbb{R}^{N\times K\times 2}$, and shifting factor, $\boldsymbol{\gamma}_i\in \mathbb{R}^{N\times K\times 2}$, are obtained using the output of distortion noise level estimator, $\hat{\mathbf{I}}_i\in \mathbb{R}^{N\times K\times 64}$.
Note that each convolution layer in denoiser block has two filters with the size of $3\times 3\times 64$ to estimate the scaling and shifting factors.
Using the obtained scaling and shifting factors, the estimated distortion noise can be obtained as 
\begin{align}
    \hat{\mathbf{Z}}_i=\boldsymbol{\alpha}_i\circ \hat{\mathbf{Y}}_i+\boldsymbol{\gamma}_i,
\end{align}
where $\circ$ denotes the hadamard product.
As a result, the estimated channel can be obtained by subtracting the pre-processed signal into the estimated distortion noise given as
\begin{align}\notag\label{prop_est_ch}
    \hat{\mathbf{H}}_i&=\hat{\mathbf{Y}}-\hat{\mathbf{Z}}_i\\
    &=(1-\boldsymbol{\alpha}_i)\circ\mathbf{H}_i+(1-\boldsymbol{\alpha}_i)\circ \mathbf{Z}_i-\boldsymbol{\gamma}_i.
\end{align}

As in \eqref{denoise_model}, the estimation error is affected by the additive distortion noise term, $\mathbf{Z}_i$. In the conventional approach of deep residual learning, the distortion noise term is extracted from the residual learning block and subtracted from the received signal. However, it cannot perfectly eliminate the distortion noise and cannot control the scale of estimated channel.
Note that the conventional LMMSE estimator reduces the MSE of estimated channel by balancing between noise enhancement and interference suppression.
As in \cite{LS_scale}, scaling of LS based estimation can be approximated as LMMSE estimator where the scaling factor can be calculated as the desired signal power divided by the received signal power.
In other words, the scaling factor assists to improve the channel estimation performance by minimizing MSE. 
Thus, the desired channel and distortion noise are scaled by $(1-\boldsymbol{\alpha}_i)$ to enhance the estimation performance.
If we set $\boldsymbol{\alpha}_i$ as zero, the proposed structure would be the same as the conventional deep residual learning.
It is also revealed that the well-trained deep residual learning is equivalent to LMMSE estimator \cite{denoiser2}.
Thus, the proposed channel estimation is also considered as LMMSE estimator when $\boldsymbol{\alpha}_i=0$ for well-trained case.
Furthermore, LMMSE estimator is no longer optimal with the existence of RHWIs but the proposed structure can provide better performance than LMMSE estimator by using the scaling factor to construct the distortion noise.
In addition, we exploit the local large-scale fading coefficients as the input of distortion noise level estimator.
Since the channel is composed of large-scale and small-scale fading and the received signal is scaled by large-scale fading, it affects to estimate the small-scale fading coefficients.
Thus, large-scale fading plays a significant role to extract the features of distortion noise.
Last but not least, proposed channel estimator does not require the knowledge of channel statistics and the level of RHWIs. Therefore, each BS can independently perform channel estimation based on its own information.
 
\subsection{Training of Channel Estimator}
In the off-line training phase, we have the training data set for the $i$-th cell as $\{\hat{\mathbf{Y}}_i^{(n)},\mathbf{1}_N\boldsymbol{\beta}_{i,i}^T, \mathbf{H}_{i,i}^{(n)}\}$ where $\hat{\mathbf{Y}}_i^{(n)}$ and $\mathbf{H}_{i,i}^{(n)}$ is the pre-processed signal for the input of the channel estimator and the ground truth at the $n$-th training sample. Also, $\boldsymbol{\beta}_{i,i}^{(n)}$ is the local large-scale fading which is known prior.
To minimize the MSE of estimated channel, the loss function of channel estimator can be written as
\begin{align}\label{c_loss}
    J_c(\boldsymbol{\Theta}_c)=\frac{1}{LN_T}\sum\limits_{n=1}^{N_T}\sum\limits_{i=1}^L\|\mathbf{H}_{i,i}^{(n)}-\hat{\mathbf{H}}_{i,i}^{(n)}\|_F^2.
\end{align}
where $\Theta_c$ is the set of parameters used in channel estimator model. 
After the training phase, the trained model is distributed to all the cells and the channel estimation performs independently in each cell at the on-line estimation phase.


\section{Joint Pilot Design and Channel Estimator}\label{sec:joint}
\subsection{Transfer Learning}
As described in \eqref{denoise_model}, the distortion noise is determined by pilot sequences even though intra-cell interference is eliminated using LS estimator.
Thus, the channel estimation performance can be further improved when the pilot sequences are well-designed.
We discussed the pilot design problem for LMMSE estimator in Sec.~\ref{sec:pilot_design}, however it is not optimized for the proposed channel estimator.
Hence, the pilot design needs to be jointly trained with the proposed channel estimator to achieve better performance whereas it takes a long training time to converge. 
For reducing the training time, we adopt the transfer learning which aims to improve the performance by exploiting the knowledge of pre-trained models.
It has the advantages of providing better initial points, boosting the performance, and accelerating the training \cite{survey_transfer}.

The proposed pilot generator is designed to minimize MSE in  \eqref{p_loss} but it requires the prior knowledge of inter-cell large-scale fading and the level of RHWIs.
Since the proposed channel estimator is designed to utilize only knowledge of local large-scale fading, the pilot generator should be also performed using the local knowledge.
Therefore, we redesign the loss function for pilot design as 
\begin{align}\label{p_loss2}
    J_p(\boldsymbol{\Theta}_p)=\frac{1}{LN_T}\sum\limits_{n=1}^{N_T}\sum\limits_{i=1}^L \mathrm{Tr}\left(\left(\mathbf{D}_{i,i}^{{(n)}^{-1}}+\frac{\tau_p}{\sigma^2}\hat{\mathbf{X}}_i^{(n)^H}\hat{\mathbf{X}}_i^{(n)}\right)^{-1}\right).
\end{align}
Note that $\bar{\mathbf{B}}_i^{(n)}$ in \eqref{p_loss} cannot be utilized due to the unknown prior knowledge.
Therefore, the loss function in \eqref{p_loss2} does not reflect the effects of pilot contamination, and thus, we use it as the initial model for joint pilot design and channel estimator. 
Since it has been known that unsupervised pre-training provides a better generalization \cite{pre_train}, pre-trained model minimizing \eqref{p_loss2} assists to the training of joint pilot design and channel estimator.
At the training phase, the pilot design model using the pre-trained model as the initial points is correlated with the channel estimator, and thus, it should be jointly trained to minimize loss in \eqref{c_loss}.
We expect that the pre-trained pilot design learns the features of LMMSE estimator which can reduce the lower bound achieved by deep residual learning based channel estimator and it also provides better performance in consideration of pilot contamination by using transfer learning.


\subsection{Complexity Analysis}
This subsection presents the computational complexity of proposed pilot design and channel estimator.
In the deep learning based pilot design, main computational operation arises from the fully-connected layer as described in Fig.~\ref{fig2}. 
In the $m$-th hidden layer with $n_m$ hidden nodes, the computation complexity is given as $\mathcal{O}(n_{m-1}n_m)$. For pilot design model, we design the number of hidden nodes, $n_m$ as $\omega\tau_pK$ where $\omega$ is the scaling parameter.
Thus, the total computation complexity of pilot design can be calculated as
\begin{align}\notag
    \mathcal{C}_\mathrm{pilot}^\mathrm{train}&= \mathcal{O}\left(N_\mathrm{ep}N_{T}L\left(n_1K+\sum \limits^{M}_{m=2} n_{m-1}n_m+2n_M\tau_pK\right)\right)\\
    &\simeq \mathcal{O}\left(N_\mathrm{ep}N_{T}L(\tau_pK)^2\right),
    \label{eq:time_coplexity_PG}
\end{align}
where $N_\mathrm{ep}$ is the number of epochs in the off-line training.
In the on-line pilot generation phase, the computation complexity in each cell can be expressed as $\mathcal{C}_\mathrm{pilot}^\mathrm{test}=\mathcal{O}\left((\tau_pK)^2\right)$.

In CNN structure used for deep residual learning, the computation complexity of the $m$-th convolution layer with the input size of $N\times K\times n_{m-1}$ is given as 
$\mathcal{O}(n_{m-1}n_mF_m^2NK)$ where $F_m$ denotes the side length of filter and $n_m$ denotes the number of output channels of the $m$-th convolutional layer.
As shown in Fig.~\ref{fig3}, LS estimator for pre-processing requires $\mathcal{O}(\tau_pN\log_2(\tau_pN))$ with inverse fast fourier transform (IFFT) implementation \cite{denoiser2}.
Moreover, deep residual learning based channel estimator is composed of $M$ convolution layers in the distortion noise level estimator and $2$ convolution layers in denoiser block.
Thus, the total computation complexity of channel estimator is calculated as
\begin{align}\notag
    {C}_\mathrm{ch}^\mathrm{train} &=  \mathcal{O}\left(N_\mathrm{ep}N_{T}NK\left(\sum\limits^{M}_{m=1}n_{m-1}n_mF_{\mathrm{N},m}^2+\sum\limits_{m'=1}^22n_{M}F_{\mathrm{D},m'}^2 \right) \right)\\
    &\simeq \mathcal{O}\left(N_\mathrm{ep}N_{T}NK\sum\limits^{M}_{m=1}n_{m-1}n_mF_{\mathrm{N},m}^2\right),
\end{align}
where $F_{\mathrm{N},m}$ and $F_{\mathrm{D},m}$ denote the side length of filter used in the distortion noise level estimator and denoiser block, respectively.
In the same way, the computational complexity in on-line estimation phase is given as $\mathcal{C}_\mathrm{ch}^\mathrm{test}= \mathcal{O}(NK\sum^{M}_{m=1}n_{m-1}n_mF_{\mathrm{N},m}^2)$.
It is note that the proposed channel estimator has the same computational complexity with the conventional deep residual learning based channel estimator in \cite{denoiser2} if it has the same number of layers.

Lastly, the joint pilot design and channel estimator utilizes the pre-trained pilot design model for transfer learning. Thus, the overall complexity of training phase is written as $\mathcal{C}_\mathrm{joint}^\mathrm{train}=\max (\mathcal{C}_\mathrm{ch}^\mathrm{train},\mathcal{C}_\mathrm{pilot}^\mathrm{train})$. Furthermore, the online prediction complexity is given by $\mathcal{C}_\mathrm{joint}^\mathrm{test}=\max(\mathcal{C}_\mathrm{ch}^\mathrm{test},\mathcal{C}_\mathrm{pilot}^\mathrm{test})$.

\section{Performance Evaluation}\label{sec:simul}
\subsection{Evaluation Settings}
\begin{figure}
    \centering
    \includegraphics[width=80mm]{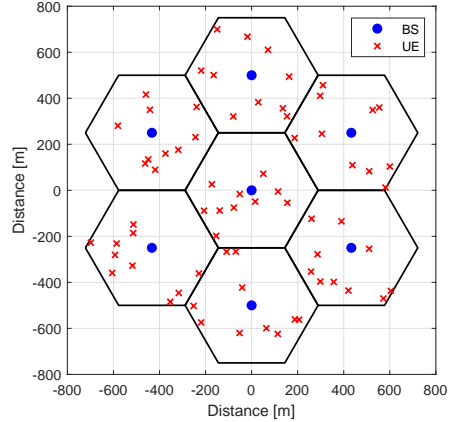}
    \caption{An illustration of hexagonal cellular network where $L=7$ and $K=10$.}
    \label{Fig1}
\end{figure}
We consider 7 hexagonal cells with 0.5\,km inter-site distance (ISD) and 10 UEs are randomly distributed in each cell with the BS at the center. It is also assumed that each UE cannot be closer than 35m to any BS as in Fig. \ref{Fig1}. Each BS is equipped with 100 antennas and each UE transmits the pilot sequence within the maximum transmit power at 23\,dBm. 
We also adopt the 3GPP LTE model \cite{LTE_standard} where the large-scale fading coefficient [dB] is modeled as
\begin{align}
    \beta_{i,j,k}=128.1+37.6\log_{10}(d_{i,j,k})+\xi_{i,j,k},
\end{align}
where $d_{i,j,k}$ is the distance in km between $\mathrm{UE}_{j,k}$ and $\mathrm{BS}_i$ and $\xi_{i,j,k}$ denotes the log-normal shadowing fading with the standard deviation 8\,dB.
Moreover, we set the noise power spectral density and spectrum bandwidth to -169\,dBm/Hz and 20\,MHz, respectively. In other words, the noise variance, $\sigma^2$, is given as -96\,dBm. For the level of RHWIs, it is assumed $\delta_{\mathrm{UE}_{i,k}}=\delta_{\mathrm{BS},i}=\delta, ~\forall~ i,k$ and we set the level of RHWIs to the range in $\delta^2\in [0, ~0.2^2]$.
For the deep learning based pilot design and channel estimator, we use Adam optimizer with the learning rate of 0.0001 and training set and test set contain 10,000 and 1,000 samples.
In addition, 2 hidden layers with $4\tau_pK$ hidden nodes are implemented for pilot design and 7 convolutional layers for the distortion noise level estimator in channel estimator are adopted.

\subsection{Evaluation Results}

Throughout this section, the proposed pilot design and channel estimator are compared with the benchmarks as follows:
\begin{itemize}
    \item \textit{Orthogonal pilot scheme}: In a fixed set of $\tau_p$ orthogonal pilot sequences, each cell randomly chooses $K$ orthogonal pilot sequences out of them.
    \item \textit{Random pilot scheme}: Pilot sequences with the length of $\tau_p$ are randomly generated in each cell. 
    \item \textit{Fractional programming (FP) pilot scheme \cite{NO_pilot3}}: To minimize the MSE of LMMSE estimator, FP approach is adopted for generating the non-orthogonal pilot sequences in each cell.
    \item \textit{CNN based deep residual network (CDRN) channel estimator \cite{denoiser2}}: The distortion noise is estimated using deep residual learning, which is subtracted from the LS estimator based pre-processed signal. It is note that the original structure of CDRN does not exploit the local large-scale fading to extract the distortion noise.
\end{itemize}



\begin{figure}
    \centering
    \includegraphics[width=90mm]{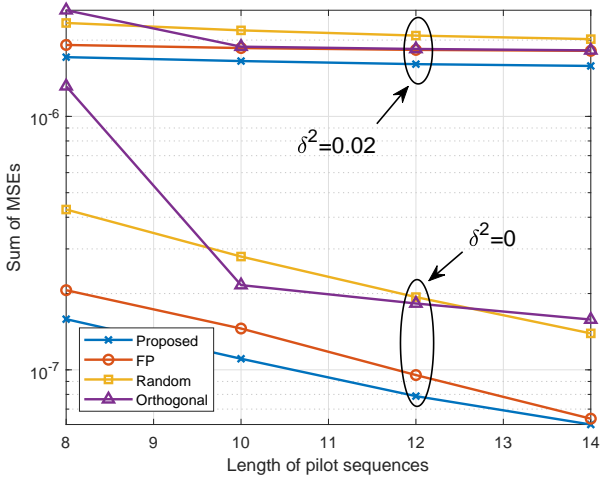}
    \caption{Sum of MSEs with respect to pilot length under the different level of RHWIs.}
    \label{simul:fig1}
\end{figure}

\begin{figure}
    \centering
    \includegraphics[width=90mm]{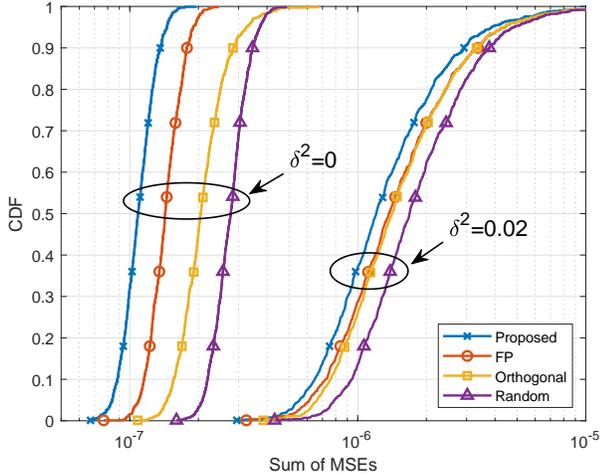}
    \caption{CDF of sum of MSEs under the different level of RHWIs where $\tau_p=10$.}
    \label{simul:fig2}
\end{figure}

Figs. \ref{simul:fig1} and \ref{simul:fig2} represent the performance of proposed pilot design with LMMSE estimator in terms of sum of MSEs at all the cells.
We notice that the proposed pilot design achieves better performance compared to FP pilot scheme.
In particular, as seen in Fig. \ref{simul:fig1}, FP exhibits almost the same MSE performance with orthogonal pilot scheme but proposed pilot design outperforms the other schemes in the existence of RHWIs.
Note that the performance of orthogonal pilot scheme is deteriorated when $\tau_p<K$ since it reuses the orthogonal pilot sequences within a cell as the limitation of pilot duration.
Thus, the other schemes offer better performance for the case of $\tau_p<K$ and the proposed pilot design exhibits considerable performance enhancement.
Furthermore, we can confirm that pilot length does not significantly affect to the estimation performance under the existence of RHWIs. Thus, it is not desirable to increase the pilot length to acquire the accurate channel information.

\begin{figure}
\centering
\subfloat[Case of zero RHWIs, $\delta^2=0$\label{simul:fig3_a}]{\includegraphics[width=90mm]{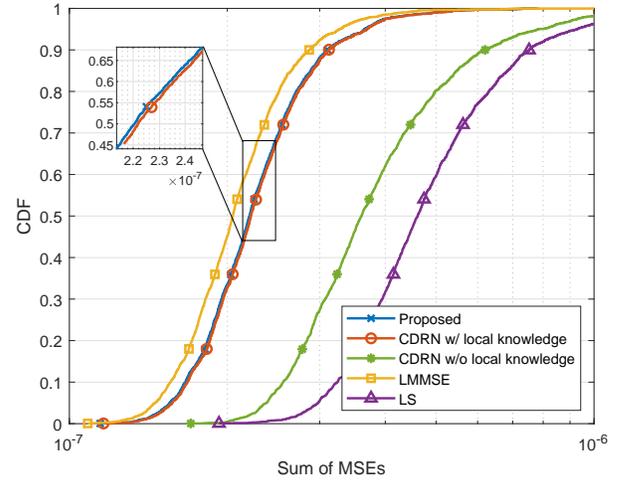}}\\
\subfloat[Case of non-zero RHWIs, $\delta^2=0.02$\label{simul:fig3_b}]{\includegraphics[width=90mm]{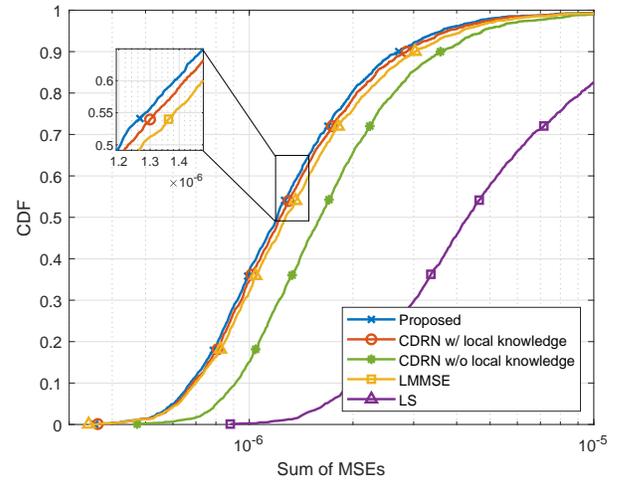}}
\caption{Comparison of sum of MSEs over different channel estimator with orthogonal pilot scheme where $\tau_p=10$.}
\label{simul:fig3}
\end{figure}

Fig. \ref{simul:fig3} compares the different channel estimators with orthogonal pilot scheme under the various level of RHWIs.
As in Fig. \ref{simul:fig3}(a), the proposed channel estimator outperforms LS estimator, which also approaches to the optimal LMMSE estimator. It implies that the distortion noise is well estimated and the effects of pilot contamination can be effectively eliminated by using deep residual learning even if the prior knowledge of pilot contamination is not exploited.
Note that CDRN with local knowledge exploits the local large-scale fading as an input of denoiser block to show the effects of local knowledge. 
We observe that the local knowledge can significantly improve the estimation performance since it assists to extract the features of distortion noise.
Furthermore, proposed channel estimator can slightly improve the estimation performance by exploiting the scaling factor over non-zero RHWIs case.
In this case, proposed channel estimator also outperforms LMMSE estimator because LMMSE estimator is no longer optimal due to the non-Gaussian distortion noise.

\begin{figure}
    \centering
    \includegraphics[width=95mm]{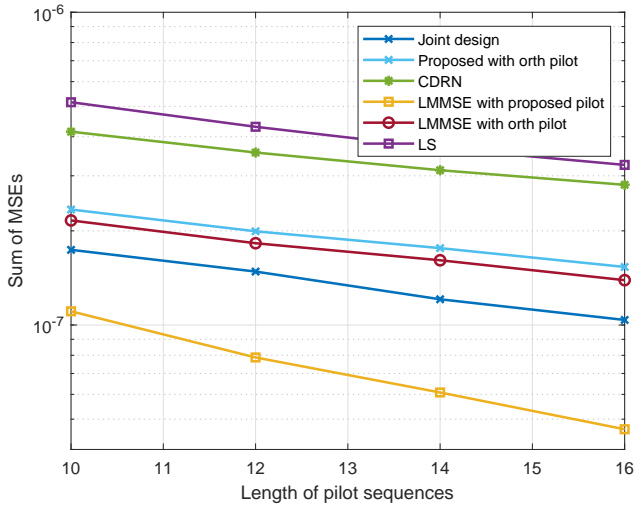}
    \caption{The sum of MSEs versus pilot length for different channel estimator with orthogonal and proposed pilot scheme under zero RHWIs.}
    \label{simul:fig4}
\end{figure}

Fig. \ref{simul:fig4} represents the sum of MSEs versus pilot length under zero RHWIs. For the pilot design, we use the orthogonal and proposed pilot schemes that is obtained by minimizing \eqref{p_loss}
to compare the proposed channel estimator with LS, LMMSE, and CDRN channel estimator.
As illustrated in Fig. \ref{simul:fig4}, proposed channel estimator with orthogonal pilot effectively eliminates the distortion noise caused by pilot contamination and the performance of it approaches to that of LMMSE estimator with orthogonal pilot even though proposed channel estimator does not exploit the prior knowledge of inter-cell large-scale fading.
More importantly, the joint pilot design and channel estimator referred to as ``Joint design" in Fig. \ref{simul:fig4} outperforms LMMSE estimator with orthogonal pilot.
It shows that pilot design plays a significant role to suppress the effects of pilot contamination. Although the aim of pre-trained pilot design model is to minimize MSE without the prior knowledge, it assists to improve the estimation performance by adopting transfer learning.
Furthermore, LMMSE with proposed pilot where  proposed pilot scheme exploits the inter-cell large-scale fading in the training phase is considered as the lower bound.
Thus, there is a gap between LMMSE with proposed pilot scheme and joint design due to the lack of prior knowledge.

\begin{figure}
    \centering
    \includegraphics[width=95mm]{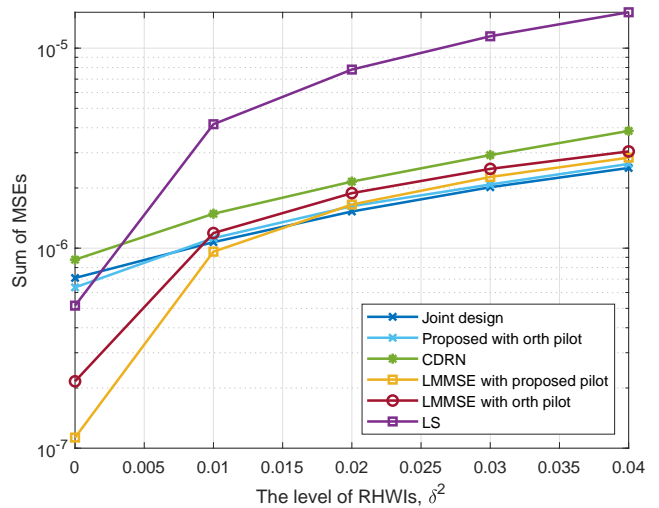}
    \caption{The sum of MSEs with the different level of RHWIs when the proposed estimator is trained with $\delta^2=0.02$ where $\tau_p=10$.}
    \label{simul:fig5}
\end{figure}

Fig. \ref{simul:fig5} illustrates the sum of MSEs with the various level of RHWIs. LMMSE estimator exploits the prior knowledge of pilot contamination while the proposed channel estimator is trained with channel samples when the level of RHWIs is $\delta^2=0.02$.
As seen in Fig. \ref{simul:fig5}, proposed channel estimator outperforms LMMSE estimator for high level of RHWIs even though the proposed channel estimator is trained with $\delta^2=0.02$.
However, the performance of proposed channel estimator with low level of RHWIs is worse than that of LMMSE estimator. Since it overestimates the distortion noise when the level of RHWIs is less than $0.02$, the desired channel information is also suppressed by denoising the distortion noise.
It is also noteworthy that the gap between the proposed pilot and orthogonal pilot schemes with LMMSE estimator is reduced as the level of RHWIs increases.
Furthermore, joint pilot design and channel estimator slightly enhance the estimation performance under non-zero RHWIs because it does not exploit the level of RHWIs for pre-trained model.
We here point out that the proposed and LMMSE channel estimators with orthogonal pilot scheme may provide enough performance when the impacts of RHWIs are dominant whereas joint design can significantly improve the estimation performance with zero RHWIs as in Fig. \ref{simul:fig4}.

\section{Conclusion}\label{sec:conclusion}
In this paper, we propose a deep residual learning based joint pilot design and channel estimator for massive MIMO systems under the consideration of hardware impairments. In particular, we exploit transfer learning for the joint training of pilot design and channel estimator to suppress the effects of pilot contamination caused by non-orthogonal pilot sequences and RHWIs. Based on the derived MSE of LMMSE estimator under the existence of RHWIs, we develop a unsupervised learning based pilot design. We also adopt a deep residual learning for channel estimator to suppress the unknown distortion noises.
Numerical simulations demonstrate and verify that the proposed joint pilot design and channel estimator effectively eliminates the effects of pilot contamination without the prior knowledge of it, which outperforms LMMSE estimator with orthogonal pilot schemes.
\section*{Appendix A}
The LMMSE estimator in \eqref{LMMSE} with non-orthogonal pilot sequences and RHWIs can be obtained as \cite{RHWI_main}
\begin{align}\label{A_1}
    \mathbf{A}_i=\arg \min \mathbb{E}\left[\|\mathbf{H}_{i,i}- \hat{\mathbf{H}}_{i,i}\|^2_\mathrm{F}\right]
\end{align}
where $\|\cdot\|_\mathrm{F}$ denotes the Frobenius norm.
Using the property of the Frobenius norm, the MSE in \eqref{A_1} can be expressed as
\begin{align}\notag\label{A_MSE}
    \mathbb{E}&\left[\|\mathbf{H}_{i,i}- \hat{\mathbf{H}}_{i,i}\|^2_\mathrm{F}\right]\\\notag
    &=\mathbb{E}\left[\mathrm{Tr}\left((\mathbf{H}_{i,i}- \mathbf{Y}_{i}\mathbf{A}_i)^H(\mathbf{H}_{i,i}- \mathbf{Y}_{i}\mathbf{A}_i)\right)\right]\\\notag
    &=\mathbb{E}\left[\mathrm{Tr}(\mathbf{H}_{i,i}^H\mathbf{H}_{i,i})\right]-\mathbb{E}\left[\mathrm{Tr}(\mathbf{A}_{i}^H\mathbf{Y}_i^H\mathbf{H}_{i,i})\right]\\
    & ~~~~-\mathbb{E}\left[\mathrm{Tr}(\mathbf{H}_{i,i}^H\mathbf{Y}_i\mathbf{A}_{i})\right]+\mathbb{E}\left[\mathrm{Tr}(\mathbf{A}_{i}^H\mathbf{Y}_i^H\mathbf{Y}_{i}\mathbf{A}_{i})\right],
\end{align}
where $\mathrm{Tr}(\cdot)$ denotes the trace operation.
By taking the derivative of \eqref{A_MSE} with respect to $\mathbf{A}_i$, the optimal LMMSE estimator can be obtained as
\begin{align}\label{A_2}
    \mathbf{A}_i=\mathbb{E}\left[\mathbf{Y}_i^H\mathbf{Y}_i\right]^{-1}\mathbb{E}[\mathbf{Y}_i^H\mathbf{H}_{i,i}].
\end{align}

In \eqref{A_2}, $\mathbb{E}\left[\mathbf{Y}_i^H\mathbf{Y}_i\right]$ can be calculated as
\begin{align}\notag\label{A_3}
    \mathbb{E}\left[\mathbf{Y}_i^H\mathbf{Y}_i\right]=&\tau_p\sum\limits_{j=1}^L\mathbb{E}\left[\mathbf{X}_j\mathbf{P}_j^{\frac{1}{2}}\mathbf{H}_{i,j}^H\mathbf{H}_{i,j}\mathbf{P}_j^{\frac{1}{2}}\mathbf{X}_j^H\right]+\mathbb{E}\left[\tilde{\mathbf{N}}_i^H\tilde{\mathbf{N}}_i\right]\\
    =&\tau_pN\sum\limits_{j=1}^L\mathbf{X}_j\mathbf{P}_j\mathbf{D}_{i,j}\mathbf{X}_j^H+\frac{1}{\tau_p}\mathbb{E}\left[\mathrm{Tr}(\mathbf{K}_{\tilde{N}_i})\right]\mathbf{I}.
\end{align}

Then, by \eqref{cov_distort}, $\mathbb{E}\left[\mathrm{Tr}(\mathbf{K}_{\tilde{N}_i})\right]$ term in \eqref{A_3}  can be written as 
\begin{align}\notag
    \mathbb{E}&\left[\mathrm{Tr}(\mathbf{K}_{\tilde{N}_i})\right]\\\notag
    &=\tau_p\sum_{j=1}^L\mathbb{E}\left[\mathrm{Tr}\left(\mathbf{H}_{i,j}(\tau_p\mathbf{\Delta}_{\mathrm{UE},j}+\delta_{\mathrm{BS},i}^2\mathbf{I})\mathbf{P}_{j}\mathbf{H}_{i,j}^H\right)\right]+\tau_pN\sigma^2\\\notag
    &=\tau_p\sum_{j=1}^L\mathrm{Tr}\left((\tau_p\mathbf{\Delta}_{\mathrm{UE},j}+\delta_{\mathrm{BS},i}^2\mathbf{I})\mathbf{P}_{j}\mathbb{E}\left[\mathbf{H}_{i,j}^H\mathbf{H}_{i,j}\right]\right)+\tau_pN\sigma^2\\\notag
    &=\tau_pN\sum_{j=1}^L\mathrm{Tr}\left((\tau_p\mathbf{\Delta}_{\mathrm{UE},j}+\delta_{\mathrm{BS},i}^2\mathbf{I})\mathbf{P}_{j}\mathbf{D}_{i,j}\right)+\tau_pN\sigma^2\\
    &=\tau_pN\sum_{j=1}^L\sum_{k=1}^K(\tau_p\delta^2_{\mathrm{UE},j,k}+\delta^2_{\mathrm{BS},i})\beta_{i,j,k}P_{j,k}+\tau_pN\sigma^2.
\end{align}

In addition, $\mathbb{E}\left[\mathbf{Y}_i^H\mathbf{H}_{i,i}\right]$ term in \eqref{A_3} can be calculated as follows using the property of independent channel between UEs, 
\begin{align}\notag\label{A_4}
    \mathbb{E}\left[\mathbf{Y}_i^H\mathbf{H}_{i,i}\right]&=\sqrt{\tau_p}\mathbf{X}_i\mathbf{P}_i^{\frac{1}{2}}\mathbb{E}\left[{\mathbf{H}_{i,i}^H\mathbf{H}_{i,i}}\right]\\\notag
    &=\sqrt{\tau_p}\mathbf{X}_i\mathbf{P}_i^{\frac{1}{2}}\mathbf{D}_{i,i}^{\frac{1}{2}}\mathbb{E}\left[{\mathbf{G}_{i,i}^H\mathbf{G}_{i,i}}\right]\mathbf{D}_{i,i}^{\frac{1}{2}}\\
    &=N\sqrt{\tau_p}\mathbf{X}_i\mathbf{P}_i^{\frac{1}{2}}\mathbf{D}_{i,i}.
\end{align}

Note that $\frac{1}{\tau_p}\mathbb{E}\left[\mathrm{Tr}(\mathbf{K}_{\tilde{N}_i})\right]$ term in \eqref{A_3} is simply expressed as $\tau_pN\phi_i$ where $\phi_i=(\sum_{j=1}^L\sum_{k=1}^K(\tau_p\delta^2_{\mathrm{UE},j,k}+\delta^2_{\mathrm{BS},i})\beta_{i,j,k}P_{j,k}+\sigma^2)/\tau_p$.
Then, the LMMSE estimator in \eqref{A_2} is obtained as
\begin{align}\label{A_5}
    \mathbf{A}_i=\frac{1}{\sqrt{\tau_p}}\left(\sum\limits_{j=1}^L\mathbf{X}_j\mathbf{D}_{i,j}\mathbf{P}_j\mathbf{X}_j^H+\phi_i\mathbf{I}\right)^{-1}\mathbf{X}_i\mathbf{P}_i^{\frac{1}{2}}\mathbf{D}_{i,i}.
\end{align}

Using the derived LMMSE estimator, the MSE of the LMMSE estimator can be represented by substituting \eqref{A_2} into \eqref{A_MSE}, therefore,
\begin{align}\notag\label{A_6}
    \mathrm{MSE}_{i}=&\mathrm{Tr}\left(\mathbb{E}\left[\mathbf{H}_{i,i}^H\mathbf{H}_{i,i}\right]\right)\\
    &-\mathrm{Tr}\left(\mathbb{E}\left[\mathbf{H}_{i,i}^H\mathbf{Y}_{i}\right]\mathbb{E}\left[\mathbf{Y}_{i}^H\mathbf{Y}_{i}\right]^{-1}\mathbb{E}\left[\mathbf{Y}_{i}^H\mathbf{H}_{i,i}\right]\right).
\end{align}

Note that in \eqref{A_6}, $\mathbb{E}\left[\mathbf{Y}_{i}^H\mathbf{Y}_{i}\right]$ and $\mathbb{E}\left[\mathbf{Y}_{i}^H\mathbf{H}_{i,i}\right]$ terms are already obtained in \eqref{A_3} and \eqref{A_4}, respectively.
As a result, the MSE can be finally derived as
\begin{align}\notag
    &\!\!\!\!\!\!\!\!\mathrm{MSE}_i\\\notag
    =&N\mathrm{Tr}(\mathbf{D}_{i,i})\\\notag
    &-N\mathrm{Tr}\Bigg(\mathbf{X}_i^H\Big(\sum\limits_{j=1}^L\mathbf{X}_j\mathbf{D}_{i,j}\mathbf{P}_j\mathbf{X}_j^H+\phi_i\mathbf{I}\Big)^{-1}\mathbf{X}_i\mathbf{P}_i\mathbf{D}_{i,i}^2\Bigg)\\\notag
    =&N\mathrm{Tr}\Bigg(\bigg(\mathbf{I}-\mathbf{X}_i^H\Big(\sum\limits_{j=1}^L\mathbf{X}_j\mathbf{D}_{i,j}\mathbf{P}_j\mathbf{X}_j^H+\phi_i\mathbf{I}\Big)^{-1}\mathbf{X}_i\mathbf{P}_i\mathbf{D}_{i,i}\bigg)\mathbf{D}_{i,i}\Bigg)\\\notag
    \overset{(a)}{=}\!&N\mathrm{Tr}\Bigg(\bigg(\mathbf{I}+\mathbf{X}_i^H\Big(\sum_{\substack{j=1\\j\neq i}}^L\mathbf{X}_j\mathbf{D}_{i,j}\mathbf{P}_j\mathbf{X}_j^H+\phi_i\mathbf{I}\Big)^{-1}\mathbf{X}_i\mathbf{P}_i\mathbf{D}_{i,i}\bigg)^{-1}\mathbf{D}_{i,i}\Bigg)\\
    =&N\mathrm{Tr}\Bigg(\bigg(\mathbf{D}_{i,i}^{-1}+\mathbf{X}_i^H\Big(\sum_{\substack{j=1\\j\neq i}}^L\mathbf{X}_j\mathbf{D}_{i,j}\mathbf{P}_j\mathbf{X}_j^H+\phi_i\mathbf{I}\Big)^{-1}\mathbf{X}_i\mathbf{P}_i\bigg)^{-1}\Bigg)
\end{align}
where $(\mathbf{A}+\mathbf{BCD})^{-1}=\mathbf{A}^{-1}-\mathbf{A}^{-1}\mathbf{B}(\mathbf{D}\mathbf{A}^{-1}\mathbf{B}+\mathbf{C}^{-1})^{-1}\mathbf{D}\mathbf{A}^{-1}$ in $(a)$.

\bibliographystyle{IEEEtran}  
\bibliography{ref_dl,ref_wj}


\end{document}